\documentclass[preprint,aps,nofootinbib,showpacs,amsfonts,epsf]{revtex4}

\input{epsf.tex}

\newcommand{\s}{\sigma}

\renewcommand{\a}{\alpha}

\newcommand{\be}{\begin{equation}}
\newcommand{\ee}{\end{equation}}
\newcommand{\bea}{\begin{eqnarray}}
\newcommand{\eea}{\end{eqnarray}}
\newcommand{\ba}{\begin{array}}
\newcommand{\ea}{\end{array}}
\def\J#1#2#3#4{{#1} {\bf #2}, #3 (#4)}
\def\PRD{Phys. Rev. D}
\def\PR{Phys. Rev.}
\def\PRL{Phys. Rev. Lett.}

\def\JMP{J. Math. Phys.}

\def\CMP{Commun. Math. Phys.}
\def\MZ{Math. Z.}
\def\TMP{Theor. Math. Phys.}

\def\JHEP{JHEP}

\def\CQG{Class. Quantum Grav.}

\def\PLA{Phys. Lett. A}
\def\PLB{Phys. Lett. B}
\def\NPB{Nucl. Phys. B}
\def\MZ{Math. Zeits.}

\begin{document}
\draft
\title{Thermodynamics of two aligned Kerr black holes}

\author{C. J. Ram\'irez-Valdez, H. Garc\'ia-Compe\'an and V.~S.~Manko}
\address{$^\dagger$Departamento de F\'\i sica, Centro de Investigaci\'on y
de Estudios Avanzados del IPN, A.P. 14-740, 07000 Ciudad de
M\'exico, Mexico}

\begin{abstract}
We investigate the first law of thermodynamics in the stationary
axisymmetric configurations composed of two Kerr black holes
separated by a massless strut. Our analysis employs the recent
results obtained for the extended double-Kerr solution and for
thermodynamics of the static single and binary black holes. We
show that, similar to the electrostatic case, in the stationary
binary systems the thermodynamic length $\ell$ is defined by the
formula $\ell=L\exp(\gamma_0)$, where $L$ is the coordinate length
of the strut, and $\gamma_0$ is the value of the metric function
$\gamma$ on the strut.
\end{abstract}


\maketitle

\section{Introduction}

In the paper \cite{AGK}, an important notion of {\it thermodynamic
length} was introduced which permits an elegant analytic
description of thermodynamics in different {\it single} black-hole
spacetimes. The usefulness of this notion has recently been
demonstrated in \cite{KZe} in application to the {\it binary}
configurations of generic charged static black holes
\cite{BMA,Man}, for which the first law of thermodynamics has been
derived in a concise form. Curiously, even in the absence of
charges, when the latter binary configurations are described by
the double-Schwarzschild solution \cite{BWe}, the work \cite{KZe}
gives for this special vacuum case a representation of the first
law different from the one considered in the well-known paper of
Costa and Perry \cite{CPe}. A natural question arises then,
whether the approach developed in the papers \cite{AGK,KZe} can be
further extended to the binary systems of {\it rotating} black
holes? In the present paper we give a positive answer to this
query.

Since the simplest rotating black hole is described by the Kerr
vacuum solution \cite{Ker}, in order to accomplish our objective
we can restrict ourselves to configurations of two Kerr black-hole
constituents kept apart in stationary equilibrium by a massless
strut \cite{Isr}. Such configurations are obtainable in principle
from the double-Kerr solution of Kramer and Neugebauer \cite{KNe}
by imposing in it the axis and asymptotic flatness conditions;
these, however, were not solved analytically in the original
parametrization of the paper \cite{KNe} without the additional
condition of the balance of sources \cite{DHo} (absence of the
strut), the fulfilment of which makes the equilibrium of two Kerr
black holes impossible \cite{MRu}. Sibgatullin's integral method
\cite{Sib} of constructing the exact solutions changed that
unpleasant situation drastically, and thanks to it we now have at
our disposal various analytical solutions for a pair of
interacting Kerr black holes separated by a massless strut which
are suitable for the study of the thermodynamic length in the
stationary binary systems. Thus, in our paper we are going to
consider the configurations of two equal counterrotating Kerr
black holes \cite{BMa,MRRS}, of two identical corotating Kerr
black holes \cite{MRu2,CCL}, and also the binary system composed
of generic Kerr black holes \cite{MRu2,MRu3,Cab}. We have decided
to analyze the configurations of equal counter- and corotating
black holes separately from the general case because the
thermodynamical properties of these particular two-body systems
were already discussed earlier in the literature \cite{HRe,HRR},
however, using exclusively the general formulas of the usual
double-Kerr solution \cite{KNe} restricted to the subextreme case
only and not elaborating the explicit form of the particular
cases; besides, the thermodynamic analysis of the corotating case
in \cite{HRR} is essentially based on numerical calculations.
Moreover, it is likely to reexamine the case of corotating black
holes because the recent paper \cite{QQS} presents an erroneous
study of such system due to employing some quantities
characteristic of exclusively the counterrotating configuration
and misinterpreting the form of the angular momentum given in
\cite{MRu2}. The simple representations of the metrics describing
the binary systems and involving the physical parameters will
allow us to obtain the concise analytic expressions for the
thermodynamic length in all the cases under consideration.

The plan of our paper is as follows. In Sec.~II we derive the
first law of thermodynamics for a pair of two equal
counterrotating black holes. Two possible ways of the derivation
of thermodynamic length are discussed. The binary system of
identical corotating Kerr black holes is considered in Sec.~III,
and the binary configurations of unequal Kerr black holes are
analyzed in Sec~IV. Concluding remarks are given in Sec.~V.

\section{Two equal counterrotating Kerr black holes}

The solution describing a binary system of equal counterrotating
Kerr sources is the vacuum specialization of the Bret\'on-Manko
electrovac solution \cite{BMa} constructed with the aid of
Sibgatullin's method, and its physical parametrization was
elaborated in the paper \cite{MRRS}, the entire metric being
defined by the line element
\be d s^2=f^{-1}[e^{2\gamma}(d\rho^2+d z^2)+\rho^2 d\varphi^2]-f(d
t-\omega d\varphi)^2, \label{Papa} \ee
with the metric coefficients $f$, $\gamma$ and $\omega$ of the
form \cite{MRRS}
\bea f&=&\frac{A\bar A-B\bar B}{(A+B)(\bar A+\bar B)}, \quad
e^{2\gamma}=\frac{A\bar A-B\bar B}{16R^4\sigma^4R_+R_-r_+r_-},
\quad \omega=-\frac{2{\rm
Im}[G(\bar A+\bar B)]}{A\bar A-B\bar B}, \nonumber\\
A&=&M^2[4\sigma^2(R_+R_-+r_+r_-)+R^2(R_+r_++R_-r_-)]
+\{(R-2M)[R(\sigma^2-a^2)+2M^3] \nonumber\\
&&+4M^2a^2\mu\}(R_+r_-+R_-r_+)-2ia\sigma R(R-2M)(R_+r_--R_-r_+),
\nonumber\\
B&=&2M\sigma R\{\sigma
R(R_++R_-+r_++r_-)-[2M^2+ia(R-2M)](R_+-R_--r_++r_-)\}, \nonumber\\
G&=&-zB+M\sigma R\{2M[2\sigma(r_+r_--R_+R_-)+R(R_-r_--R_+r_+)]
\nonumber\\ &&+(R+2\sigma)[R\sigma-2M^2-ia(R-2M)](R_+-r_-)
\nonumber\\
&&+(R-2\sigma)[R\sigma+2M^2+ia(R-2M)](R_--r_+)\}, \nonumber\\
R_\pm&=&\sqrt{\rho^2+(z+{\textstyle\frac{1}{2}}R\pm\sigma)^2},
\quad r_\pm=\sqrt{\rho^2+(z-{\textstyle\frac{1}{2}}R\pm\sigma)^2},
\label{metric1} \eea
where $M$ is the mass of each black hole, $a$ is the angular
momentum per unit mass of the lower black hole ($-a$ for the upper
black hole, see Fig.~1), $R$ is the coordinate distance between
the centers of black holes, while the constant quantity $\s$
representing the half length of each black hole's horizon is given
by the formula
\be \sigma=\sqrt{M^2-a^2\mu}, \quad \mu\equiv\frac{R-2M}{R+2M}.
\label{sigma1} \ee
We would like to emphasize that the metric
(\ref{Papa})-(\ref{sigma1}) describes the configurations of two
counterrotating black holes or hyperextreme sources. However,
since our interest lies in the black-hole sector of the above
solution, the parameters $M$, $a$ and $R$ must preserve the
reality of $\s$, which implies $\s^2>0$.

It was shown in \cite{MRRS} that each black hole in the solution
(\ref{Papa})-(\ref{sigma1}) satisfies the well-known Smarr mass
formula \cite{Sma}
\be M=\frac{1}{4\pi}\kappa{\cal A}+2\Omega J, \label{Smar} \ee
where $\kappa$ is the surface gravity, ${\cal A}$ the area of the
horizon, $\Omega$ the lower black hole horizon's angular velocity,
and $J$ the angular momentum of the lower black hole ($-\Omega$
and $-J$ in the case of the upper black hole). Then $J=Ma$, while
for ${\cal A}$, $\kappa$ and $\Omega$ the paper \cite{MRRS} gives
the expressions\footnote{Note that in \cite{MRRS} the letter $S$
was used for denoting horizon's area, but in our paper $S$ stands
for the entropy.}
\be {\cal A}=8\pi M(M+\sigma)\Bigl(1+\frac{2M}{R}\Bigr), \quad
\kappa=\frac{R\s}{2M(M+\sigma)(R+2M)}, \quad
\Omega=\frac{a\mu}{2M(M+\s)}. \label{area1} \ee

Formulas (\ref{area1}) together with the expression of the
interaction force \cite{Var}
\be {\cal F}=\frac{M^2}{R^2-4M^2} \label{F1} \ee
permit us to elaborate the first law of thermodynamics for the
binary system under consideration by following the procedure
similar to the one employed in the paper \cite{KZe}. Passing from
the area ${\cal A}$ to the entropy $S$ via
$S=\textstyle{\frac{1}{4}}{\cal A}$ \cite{Bek,Haw}, we must take
differentials of the quantities $S$, $\Omega$ and ${\cal F}$ by
considering these as functions of the parameters $(M,a,R)$ or
parameters $(M,J,R)$. The second option seems more simple, and to
use it, one only has to change $a$ to $J/M$ in the expressions of
$\Omega$ and $\s$. After having obtained the form of $dS$,
$d\Omega$ and $d{\cal F}$ in terms of $dM$, $dJ$, $dR$, one has to
solve the resulting algebraic system for $dM$, $d\Omega$ and $dR$,
thus yielding the desired expression for $dM$ in terms of $dS$,
$dJ$ and $d{\cal F}$:
\be dM=\frac{R\s }{4\pi M(R+2M)(M+\sigma)}\,dS+\frac{J\mu }{2
M^2(M+\s)}\,dJ-\frac{(R-2\s)(R^2-4M^2)}{2R^2}\,d{\cal F},
\label{dM1} \ee
so that we can introduce the temperature $T$ by
\be T=\left.\frac{\partial M}{\partial S}\right|_{J,{\cal F}}
=\frac{R\s }{4\pi M(R+2M)(M+\sigma)}, \label{T1} \ee
and from (\ref{area1}) it follows that the above $T$ coincides
with the Hawking temperature $T=\kappa/(2\pi)$ \cite{Haw}.
Therefore, taking into account the equality of black holes, we
finally arrive at the first law of thermodynamics for the entire
system in the form
\bea dM_T&=&2TdS+2\Omega dJ-\ell d{\cal F}, \nonumber\\
M_T&=&2M, \quad \ell=(R-2\s)(R^2-4M^2)/R^2. \label{L1} \eea
The last term on the right-hand side in (\ref{L1}) determines the
contribution of the conical singularity into the first law, and it
represents elementary work performed by the strut. It can be
easily seen that the thermodynamic length $\ell$ reduces in the
static limit to the respective $\ell$  for two equal Schwarzschild
black holes (see formula (4.5) of \cite{KZe} in the case $m=M$).

By observing that $R-2\s$ is the coordinate length of the strut,
and $(R^2-4M^2)/R^2$ is equal to $\exp(\gamma_0)$, $\gamma_0$
being the value of the metric function $\gamma$ on the strut (the
part $-\textstyle{\frac{1}{2}}R+\s\le
z\le\textstyle{\frac{1}{2}}R-\s$ of the symmetry axis), we arrive
at the remarkable conclusion that $\ell$ is defined by the same
formula as obtained in \cite{KZe} for the static case:
\be \ell=Le^{\gamma_0}, \label{lL} \ee
where $L=R-2\s$. Of course, this could be a mere coincidence which
might be attributed to the same form of the interaction force
(\ref{F1}) as in the case of two equal Schwarzschild black holes.
So, further evidence is still needed to make sure that (\ref{lL})
holds generically for other stationary binary systems of black
holes too.

To conclude this section, we would like to remark that the
derivation of the first law (\ref{L1}) can be also performed
directly using the parameter set $(M,a,R)$ which enters the
formulas (\ref{metric1})-(\ref{sigma1}). This possibility, as will
be seen later, is highly important when the rotational parameter
$a$ is related in a complicated way to the angular momentum $J$.
Since in our case $J$ is simply $Ma$ and hence $dJ=Mda+adM$, one
will be able to arrive at the correct result for $dM$ in
(\ref{dM1}) by substituting the differentials $da$ by $(dJ-adM)/M$
throughout the calculations.

\section{Two equal corotating Kerr black holes}

The case of two identical corotating Kerr black holes is described
by the exact solution worked out in two different representations
in the papers \cite{MRu2,CCL}. The representation involving
physical parameters, which is of interest to us for our purposes,
is defined by the formulas
\bea f&=&\frac{A\bar A-B\bar B}{(A+B)(\bar A+\bar B)}, \quad
e^{2\gamma}=\frac{A\bar A-B\bar B}{K_0^2R_+R_-r_+r_-}, \quad
\omega=4a-\frac{2{\rm Im}[G(\bar A+\bar B)]}{A\bar A-B\bar B},
\nonumber\\ A&=&R^2(R_+-R_-)(r_+-r_-)
-4\s^2(R_+-r_+)(R_--r_-), \nonumber\\
B&=&2R\s[(R+2\s)(R_--r_+)-(R-2\s)(R_+-r_-)], \nonumber\\
G&=&-zB +R\s[2R(R_-r_--R_+r_+) +4\s(R_+R_--r_+r_-)\nonumber\\
&&-(R^2-4\s^2)(R_+-R_--r_++r_-)], \nonumber\\
R_\pm&=&\frac{-M(\pm2\s+R)+id}{2M^2+(R+2ia)(\pm\s+ia)}
\sqrt{\rho^2+\left(z+\frac{1}{2}R\pm\s\right)^2}, \nonumber\\
r_\pm&=&\frac{-M(\pm2\s-R)+id}{2M^2-(R-2ia)(\pm\s+ia)}
\sqrt{\rho^2+\left(z-\frac{1}{2}R\pm\s\right)^2}, \nonumber\\
K_0&=&\frac{4\s^2[(R^2+2MR+4a^2)^2
-16M^2a^2]}{M^2[(R+2M)^2+4a^2]}, \label{metric2} \eea
where the constant quantities $\s$ and $d$ have the form
\be \s=\sqrt{M^2-a^2+d^2 (R^2-4M^2+4a^2)^{-1}}, \quad
d=\frac{2Ma(R^2-4M^2+4a^2)} {R^2+2MR+4a^2}. \label{sigma2} \ee
Like in the previous case of counterrotating black holes, the
parameters $M$ and $R$ denote, respectively, the mass of each
black hole and the coordinate distance between the centers of
black holes (see Fig.~2); however, now the rotational parameter
$a$ is not equal exactly to the angular momentum $J$ of the black
hole per unit mass $M$, but its relation to $J$ is determined by
the following cubic equation:
\be J=\frac{Ma[(R+2M)^2+4a^2]}{R^2+2MR+4a^2}. \label{Ja2} \ee
The black-hole sector of the metric (\ref{metric2})-(\ref{sigma2})
corresponds to the real-valued $\s$, while the hyperextreme Kerr
sources, which are of no interest to us in this paper, are
described by the pure imaginary $\s$.

Each black hole in the binary system verifies identically the
Smarr formula (\ref{Smar}), and the known thermodynamical
characteristics which we will need for the derivation of the first
law of thermodynamics are written down below:
\bea S&=&\frac{2\pi M[(R+2M)^2+4a^2][(R+2M)(M+\s)-2a^2]}
{(R+2\sigma)(R^2+2MR+4a^2)}, \nonumber\\
T&=&\frac{\s(R+2\sigma)(R^2+2MR+4a^2)} {4\pi
M[(R+2M)^2+4a^2][(R+2M)(M+\s)-2a^2]},
\nonumber\\\Omega&=&\frac{(M-\sigma)
(R^2+2MR+4a^2)}{2Ma[(R+2M)^2+4a^2]}, \nonumber\\ {\cal F}&=&
\frac{M^2[(R+2M)^2-4a^2]}{(R^2-4M^2+4a^2)[(R+2M)^2+4a^2]},
\label{ST2} \eea
where we have given the expression of the entropy $S$ instead of
the horizon area ${\cal A}$,\footnote{There is a misprint in the
formula (26) of \cite{MRu2} for horizon's area: the last term in
the numerator must read $-2a^2$.} and the temperature $T$ instead
of the surface gravity $\kappa$.

The parameter set that we must employ during the calculations is
$(M,a,R)$ which does not include explicitly the angular momentum
$J$. Therefore, we have to follow the procedure outlined at the
end of the previous section, i.e., we should treat the quantities
$S$, $\Omega$ and ${\cal F}$ as functions of $M$, $a$ and $R$, and
after taking differentials $dS$, $d\Omega$ and $d{\cal F}$ we must
change the differential $da$ to the combination of the
differentials $dJ$, $dM$ and $dR$ via the formula obtainable from
(\ref{Ja2}), namely,
\bea da&=&\frac{1}{M[R(R+2M)^3+8a^2(R^2+MR-2M^2+2a^2)]}
\{(R^2+2MR+4a^2)^2\,dJ \nonumber\\
&&-a[(R+2M)^2(R^2+4MR+8a^2)+16a^2(m^2+a^2)]\,dM \nonumber\\
&&+2M^2a[(R+2M^2)^2-4a^2]\,dR\}. \label{da} \eea
Then it only remains to solve the system of three algebraic
equations for $dM$, $d\Omega$ and $dR$, and the expression for
$dM$ multiplied by 2 (due to equality of black holes) finally
provides us with the first law of thermodynamics for the binary
configuration of two identical corotating Kerr black holes:
\bea dM_T&=&2TdS+2\Omega dJ-\ell d{\cal F}, \nonumber\\
M_T&=&2M, \quad \ell=\frac{(R-2\s)(R^2-4M^2+4a^2)[(R+2M)^2+4a^2]}
{(R^2+2MR+4a^2)^2-16M^2a^2}, \label{L2} \eea
with the coefficients $T$ and $\Omega$ defined by (\ref{ST2}). One
can see that the first law (\ref{L2}) has the same structure as in
(\ref{L1}). Although the thermodynamic length $\ell$ in (\ref{L2})
has a more complicated form than the respective $\ell$ in
(\ref{L1}), it is still not difficult to verify that the new
$\ell$ obeys formula (\ref{lL}) too: the coordinate length $L$ of
the strut is the same as in the previous `counterrotating' case
($L=R-2\s$) and the value of $\exp(\gamma_0)$ calculated with the
aid of formulas (\ref{metric2})-(\ref{sigma2}) coincides with
$\ell/L$ in (\ref{L2}).

We now turn to the general case of rotating Kerr black holes.

\section{Two generic Kerr black holes}

The general solution describing a system of two aligned Kerr black
holes separated by a massless strut is defined by the formulas
\cite{MRu2,MRu3,Cab}
\bea f&=&\frac{A\bar A-B\bar B}{(A+B)(\bar A+\bar B)}, \quad
e^{2\gamma}=\frac{A\bar A-B\bar B}{16|\s_1|^2|\s_2|^2K_0^2 \tilde
R_+\tilde R_-\tilde r_+\tilde r_-}, \quad \omega=2a-\frac{2{\rm
Im}[G(\bar A+\bar B)]}{A\bar A-B\bar B}, \nonumber\\
A&=&[R^2-(\s_1+\s_2)^2](R_+-R_-)(r_+-r_-)
-4\s_1\s_2(R_+-r_-)(R_--r_+), \nonumber\\
B&=&2\s_1(R^2-\s_1^2+\s_2^2)(R_--R_+)
+2\s_2(R^2+\s_1^2-\s_2^2)(r_--r_+) \nonumber\\
&&+4R\s_1\s_2(R_++R_--r_+-r_-), \nonumber\\
G&=&-zB +\s_1(R^2-\s_1^2+\s_2^2)(R_--R_+)(r_++r_-+R) \nonumber\\
&&+\s_2(R^2+\s_1^2-\s_2^2)(r_--r_+)(R_++R_--R) \nonumber\\
&&-2\s_1\s_2\{2R[r_+r_--R_+R_--\s_1(r_--r_+)+\s_2(R_--R_+)]
\nonumber\\ &&+(\s_1^2-\s_2^2)(r_++r_--R_+-R_-)\}, \nonumber\\
r_\pm&=&\mu_0^{-1}\frac{(\pm\s_1-m_1-ia_1)[(R+M)^2+a^2]
+2a_1[m_1a+iM(R+M)]} {(\pm\s_1-m_1+ia_1)[(R+M)^2+a^2]
+2a_1[m_1a-iM(R+M)]}\,\tilde r_\pm, \nonumber\\
R_\pm&=&-\mu_0\frac{(\pm\s_2+m_2-ia_2)[(R+M)^2+a^2]
-2a_2[m_2a-iM(R+M)]} {(\pm\s_2+m_2+ia_2)[(R+M)^2+a^2]
-2a_2[m_2a+iM(R+M)]}\,\tilde R_\pm, \nonumber\\
\tilde r_\pm&=&\sqrt{\rho^2+
\left(z-\frac{1}{2}R\pm\s_1\right)^2}, \quad \tilde
R_\pm=\sqrt{\rho^2+ \left(z+\frac{1}{2}R\pm\s_2\right)^2},
\label{metric3} \eea
where the constants $K_0$ and $\mu_0$ have the form\footnote{We
have rectified misprints in the formula (13) of \cite{MRu3} for
$K_0$ and formulas (5) of \cite{MRu3} for $d_1$ and $d_2$.}
\be K_0=\frac{[(R+M)^2+a^2] [R^2-(m_1-m_2)^2+a^2]-8m_1m_2a^2}
{m_1m_2[(R+M)^2+a^2]}, \quad \mu_0=\frac{R+M-ia}{R+M+ia},
\label{K0} \ee
and the quantities $\s_1$ and $\s_2$ representing the half lengths
of the horizons of black holes are given by the expressions
\bea \s_1=\sqrt{m_1^2-a_1^2+4m_2a_1d_1}, \quad
\s_2=\sqrt{m_2^2-a_2^2+4m_1a_2d_2}, \nonumber\\ d_1=\frac{
[m_1(a_1-a_2+a)+Ra_1][(R+M)^2+a^2]+m_2a_1a^2} {[(R+M)^2+a^2]^2},
\nonumber\\ d_2=\frac{ [m_2(a_2-a_1+a)+Ra_2][(R+M)^2+a^2]
+m_1a_2a^2}{[(R+M)^2+a^2]^2}. \label{s1s2} \eea

The arbitrary real parameters of the metric
(\ref{metric3})-(\ref{s1s2}) are $m_1,m_2,a_1,a_2$ and $R$, five
in total, and the upper black hole has mass $m_1$ and angular
momentum per unit mass $a_1$, while the lower black hole is
endowed with mass $m_2$ and angular momentum per unit mass $a_2$,
so that $a_1=j_1/m_1$, $a_2=j_2/m_2$, $j_1$ and $j_2$ being
angular momenta of the upper and lower black hole, respectively
(see Fig.~3); we note that these masses and angular momenta are
Komar quantities \cite{Kom}. As usual, the parameter $R$ denotes
the coordinate distance between the centers of black holes. The
total mass $M$ and total angular momentum $J$ of the binary system
have the form
\be M=m_1+m_2, \quad J=m_1a_1+m_2a_2, \label{MJ} \ee
and $a$ is related to the aforementioned five parameters by the
cubic equation
\be (R^2-M^2+a^2)(a_1+a_2-a)+2(R+M)(J-Ma)=0, \label{cubic} \ee
so that its role is similar to that of $a$ from the previous
section. Note that the particular case of equal counterrotating
black holes follows from the general formulas by setting
$m_1=m_2=M$, $a_2=-a_1=\a$, $a=0$, while the corotating case of
equal black holes corresponds to the parameter choice $m_1=m_2=M$,
$a_1=a_2=\a$, with a formal redefinition $a\to 2a$ and changing
$\a$ to $a$ by means of (\ref{cubic}).

For the entropies $(S_1,S_2)$, temperatures $(T_1,T_2)$, horizon's
angular velocities $(\Omega_1,\Omega_2)$ of each black hole, and
the interaction force ${\cal F}$ we have the expressions
\bea \frac{S_1}{\pi}&=&\frac{\sigma_1}{2\pi T_1}=\frac{
\{(m_1+\s_1)[(R+M)^2+a^2]-2m_1a_1a\}^2 +a_1^2(R^2-M^2+a^2)^2}
{[(R+M)^2+a^2][(R+\s_1)^2-\s_2^2]}, \nonumber\\
\frac{S_2}{\pi}&=&\frac{\sigma_2}{2\pi T_2}=\frac{
\{(m_2+\s_2)[(R+M)^2+a^2]-2m_2a_2a\}^2 +a_2^2(R^2-M^2+a^2)^2}
{[(R+M)^2+a^2][(R+\s_2)^2-\s_1^2]}, \nonumber\\
\Omega_1&=&\frac{m_1-\s_1}{2m_1a_1}, \quad
\Omega_2=\frac{m_2-\s_2}{2m_2a_2}, \nonumber\\ {\cal F}&=&
\frac{m_1m_2[(R+M)^2-a^2]}{(R^2-M^2+a^2)[(R+M)^2+a^2]},
\label{S12} \eea
that must now be cleverly used for the derivation of the first law
of thermodynamics. To avoid the resolution of the cubic equation
(\ref{cubic}), it appears that the best strategy to tackle the
derivation problem is to work with the parameter set
$\{m_1,m_2,j_1,R,a\}$, for which purpose it is necessary first to
change $a_1$ and $a_2$ to $j_1/m_1$ and $j_2/m_2$ in the formulas
(\ref{S12}), as well as in the expressions for $\s_1$ and $\s_2$.
Then we have to solve equation (\ref{cubic}) for $j_2$, yielding
\be j_2=\frac{m_2\{m_1a[(R+M)^2+a^2]-j_1[(R+m_1)^2-m_2^2+a^2]\}}
{m_1[(R+m_2)^2-m_1^2+a^2]}, \label{jj2} \ee
and make another substitution in the formulas involved, this time
changing $j_2$ by means of (\ref{jj2}). As a result, we have the
necessary formulas rewritten in the desired parameter set and can
proceed in a standard way. We must take differentials of the
quantities $S_1$, $S_2$, $\Omega_1$, $\Omega_2$ and ${\cal F}$ by
considering these as functions of $m_1,m_2,j_1,R,a$ and changing
the differentials $da$ to $dj_2$ by means of (\ref{jj2}). The
resulting system of five algebraic equations must be solved for
$dm_1$, $dm_2$, $d\Omega_1$, $d\Omega_2$ and $dR$, thus giving us
the first law of thermodynamics as the sum of $dm_1$ and $dm_2$:
\bea dM&=&T_1\,dS_1+T_2\,dS_2+\Omega_1 \,dj_1+\Omega_2 \,dj_2
-\ell \,d{\cal F}, \nonumber\\
M&=&m_1+m_2, \quad
\ell=\frac{(R-\s_1-\s_2)(R^2-M^2+a^2)[(R+M)^2+a^2]}
{(R^2+MR+a^2)^2-(m_1-m_2)^2(R+M)^2-4m_1m_2a^2}. \label{L3} \eea
Therefore, as it follows from (\ref{L3}), the structure of the
thermodynamic length $\ell$ in the general case remains the same
as in two previous particular cases -- it is determined by the
formula (\ref{lL}) because the coordinate length of the strut for
the unequal black holes is $L=R-\s_1-\s_2$, and the part $\ell/L$
in (\ref{L3}) coincides exactly with the corresponding value of
$\exp(\gamma_0)$, as can be easily verified. It is quite
surprising that the very cumbersome intermediate calculations have
eventually led us to an elegant final result for $\ell$ proving
the universal character of the formula (\ref{lL}).

\section{Concluding remarks}

In the present paper we have succeeded in extending the notion of
thermodynamic length $\ell$ further to binary stationary systems
of black holes and found the explicit concise form of $\ell$ in
the case of three different binary configurations. The
thermodynamic length permits one to derive analytically the first
law of thermodynamics in a consistent way, and the physical
parametrization of the solutions describing the systems of black
holes simplifies considerably the derivation procedure. It is
remarkable that $\ell$ in the stationary case turns out to be
determined by the same formula (\ref{lL}) as in the static vacuum
and electrostatic cases, and it admits the same geometrical
interpretation as given in \cite{KZe} -- the area of the
worldsheet of the strut per unit time. This suggests in particular
that most probably the notion of thermodynamic length is also
applicable to the stationary electrovac configurations of black
holes as well. At least this is certainly true in the case of the
Bret\'on-Manko solution \cite{BMa} for two equal counterrotating
Kerr-Newman black holes \cite{New}. Using the physical
parametrization of this solution obtained in \cite{MRR} it can be
actually shown that the corresponding first law of thermodynamics
and the thermodynamic length have the form
\bea dM_T&=&2TdS+2\Omega dJ+2\Phi dQ-\ell d{\cal F}, \nonumber\\
M_T&=&2M, \quad \ell=(R-2\s)(R^2-4M^2+4Q^2)/R^2, \label{L4} \eea
($Q$ is the charge and $\Phi$ the electric potential) and $\ell$
verifies formula (\ref{lL}). In the absence of rotation, one
recovers the result obtained for $\ell$ in \cite{KZe}. We are
going to consider the thermodynamics of rotating charged binary
black holes in a separate publication.

\section*{Acknowledgments}

We are thankful to the anonymous referee for valuable suggestions.
This work was supported in part by Project~128761 from CONACyT of
Mexico.

\begin{figure}[htb]
\centerline{\epsfysize=75mm\epsffile{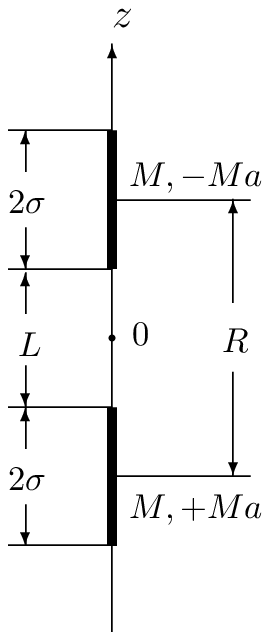}} \caption{Location
of two equal conterrotating Kerr black holes on the symmetry
axis.}
\end{figure}

\begin{figure}[htb]
\centerline{\epsfysize=75mm\epsffile{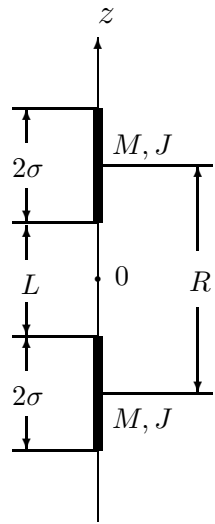}} \caption{Location
of two identical corotating Kerr black holes on the symmetry
axis.}
\end{figure}

\begin{figure}[htb]
\centerline{\epsfysize=75mm\epsffile{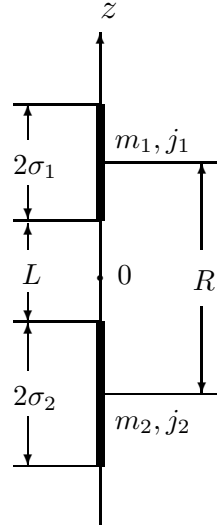}} \caption{Location
of two unequal Kerr black holes on the symmetry axis. The
coordinate length of the strut $L$ is equal to $R-\s_1-\s_2$.}
\end{figure}

\end{document}